\documentclass[aps,prl,preprint]{revtex4}
\usepackage{graphicx}

\begin{document}

\title{Noise-spectroscopy of multiqubit systems: Determining all their parameters by applying an external classical noise}

\author{S. Savel'ev$^{1,2}$, A.M. Zagoskin$^{1,2}$, A.N. Omelyanchouk$^{2,3}$, Franco Nori$^{2,4}$}
\affiliation{$^1$Department of Physics, Loughborough University, Leicestershire, LE11 3TU, United Kingdom\\$^2$Advanced Science Institute,
RIKEN, Wako-shi, Saitama 351-0198, Japan\\ 
$^3$B.Verkin Institute for Low Temperature Physics and Engineering,  61103, Kharkov, Ukraine\\
$^4$Physics Department,
The University of Michigan, Ann Arbor, MI 48109-1040, USA}

\begin{abstract}
Imagine that you have several sets of two coupled qubits, but you do not know the parameters of their Hamitonians. How to determine these without resorting to the usual 
spectroscopy 
approach to the problem? 
Based on numerical modeling, we show that all the parameters of a system of two coupled qubits can be determined by applying to it an external classical noise and analysing the Fourier spectrum of the elements of the system's density matrix. In particular, the interlevel spacings as well as the strength and sign of qubit-qubit coupling can be determined this way. 
\end{abstract}
\maketitle

\section{introduction}

Despite steady successes in fabrication and measurement techniques, the experimental characterization of multi-qubit systems \cite{PhysTod,PhysCan} remains a challenge due to their complicated level structure.  Our goal here is to determine the system's {\em parameters}, as distinct from the more difficult problem of determining its {\em state}, which has to be tackled using quantum state tomography \cite{QST,QST-exp}.  For example, neither the strength nor the sign of the qubit-qubit  coupling are known a priori. One of several standard approaches   studies the resonant response  of quantum macroscopic systems to an external coherent signal (see, e.g., \cite{i1,i2,delft1}),  allowing to determine the qubit parameters by scanning the frequency range of the external signal. The difficulty in the straightforward application of this approach, due to the fact that only few qubits can be actually accessed, and the relation of this problem to the general field of inverse problems, were addressed in \cite{burg1,burg2}.

An alternative approach to the standard spectroscopic  methods  of characterization would  use as a drive a broad-band noise. We call it {\em active noise spectroscopy}, as distinct from the ``passive" noise spectroscopy of Ref.~\cite{i1}, where the response of the noise spectrum to a {\em coherent monochromatic} drive was measured.   

Recently, we have shown \cite{om} that classical noise applied to a qubit produces persistent oscillations of the  off-diagonal density matrix elements (``coherences") despite finite dephasing and relaxation times. In other words, a moderate amount of external noise {\em enhances} quantum coherence, which manifests in oscillations with a frequency corresponding to quantum transitions between the ground and first excited states. There exists an optimal noise amplitude: at lower noise level,  oscillations are suppressed, while as the noise is increased, the oscillations become random and the corresponding spectroscopic peak is eventually smeared away. 
 Indeed, for zero noise, the oscillations of the off-diagonal elements of the density matrix decay on the time scale of $\tau$, where $1/\tau$ is the dephasing rate. Moderate phase-insensitive noise excites the system from time to time, allowing the qubit to evolve with its own frequency between the relatively rare noise spikes, thus, uncovering quantum dynamics. Strong noise produces strong spikes very often, thus leaving no time for the coherent evolution. This phenomenon is related to both classical and quantum stochastic resonances, which manifest in various physical systems (see, e.g., \cite{han1,han2,han3,han4,hue1,Galve}).                 

In this paper we investigate  how these effects of classical noise can help determine the parameters of a multiqubit system. Specifically, we consider two coupled qubits and analyze the spectrum of the density matrix excited by white Gaussian classical noise. We numerically show that the resulting noise spectra contain four peaks, which  correspond to the interlevel transitions in the system. From these, the energy spectrum and all the model parameters of the qubits are readily obtained. In addition, the correlations in the matrix elements corresponding to different qubits can be used to conclude whether the qubits are coupled ferro- or antiferromagnetically.

\section{Model}

Two coupled qubits can be  described by the Hamiltonian \cite{app}
\begin{equation}
	H = -\frac{1}{2} \sum_{j=1,2} \left[\Delta_j \sigma^j_z + \epsilon_j(t)\sigma^j_x\right] + g \sigma^1_x\sigma^2_x
	\label{eq-ham}
\end{equation}
where $\sigma^j_z$ and $\sigma^j_x$ are Pauli matrices corresponding to either the first ($j=1$) or the second ($j=2$) qubits, and the eigenstates of $\sigma^j_z$
are the basis states in the localized representation of the $j$th qubit at zero coupling.  Note that the results obtained below do not 
qualitatively depend on the type of coupling (e.g., $\sigma_x^1\sigma_x^2$ versus $\sigma_y^1\sigma_y^2$): in any case the noise will allow to determine
the parameters of the two-qubit Hamiltonian. For this reason, and for demonstrating the physical principles of noise-induced spectroscopy, we consider two identical qubits. The tunneling
splitting energies $\Delta_{1,2}$ (in case of the identical qubits we will be investigating here: $\Delta_1=\Delta_2=\Delta$) are determined by the design and fabrication details of the  device, while the bias energies $\epsilon_j(t)$ can be controlled externally and, in our case, are only driven by the noise, 
\begin{equation}
\epsilon_j(t) = \delta\!\xi_j(t).
\end{equation}
The Gaussian white noise considered here is zero-averaged and delta-correlated:
\begin{equation}
\langle\delta\! \xi_j(t)\rangle = 0,\:\: 	\langle\delta\! \xi_j(t)\delta\! \xi_{j'}(t)\rangle = 2D\delta_{j,j'}\delta(t-t').
\label{noise}
\end{equation}
 where $D$ is the noise intensity, which should be defined for each particular system (see, e.g., the example of two flux qubits 
described below). The uncorrelated noise sources affecting the qubits (``local'' noise) tend to be more detrimental to their quantum coherence than the correlated ones \cite{Storcz,You,wilh,doll,Zhang}, which makes Eq.~(\ref{noise}) the ``worst case scenario''.

\subsection{Master equation}

By writing the qubit density matrix $\hat{\rho}$ as 
\begin{equation}
\hat{\rho}=\frac{1}{4} \sum_{a,b=0,x,y,z}\Pi_{ab}
\; \sigma^1_a \otimes \sigma^2_b
\end{equation}
we can rewrite the master equation 
$$\frac{d\hat{\rho}}{dt} = -i\left[\hat{H}(t),\hat{\rho}\right]+\hat{\Gamma}\hat{\rho}$$ 
in the form 
\begin{eqnarray}
\begin{array}{lll}
\dot{\Pi}_{0x}	& = & \Delta_2\Pi_{0y} - \Gamma_{\phi 2}\Pi_{0x} \\
\dot{\Pi}_{0y}	& = & -\Delta_2\Pi_{0x} + \epsilon_2(t)\Pi_{0z} - 2g\Pi_{xz} - \Gamma_{\phi 2}\Pi_{0y}\\
\dot{\Pi}_{0z}	& = & -\epsilon_2(t)\Pi_{0y} + 2g\Pi_{xy}-  \Gamma_{2}(\Pi_{0z}-Z_{T2})\\
& & \\
\dot{\Pi}_{x0}	& = & \Delta_1\Pi_{y0} - \Gamma_{\phi 1}\Pi_{x0} \\
\dot{\Pi}_{y0}	& = & -\Delta_1\Pi_{x0} + \epsilon_1(t)\Pi_{z0} - 2g\Pi_{zx} - \Gamma_{\phi 1}\Pi_{y0}\\
\dot{\Pi}_{z0}	& = & -\epsilon_1(t)\Pi_{y0} + 2g\Pi_{yx}-  \Gamma_{1}(\Pi_{z0}-Z_{T1})\\
& & \\
\dot{\Pi}_{xx}	& = & \Delta_2\Pi_{xy} + \Delta_1\Pi_{yx} - (\Gamma_{\phi 1} + \Gamma_{\phi 2})\Pi_{xx} \\
& & \\
\dot{\Pi}_{xy}	& = & -2g\Pi_{0z} -\Delta_2\Pi_{xx} + \Delta_1\Pi_{yy} + \epsilon_2(t)\Pi_{xz} -    (\Gamma_{\phi 1} + \Gamma_{\phi 2})\Pi_{xy}\\
\dot{\Pi}_{yx}	& = & -2g\Pi_{z0} -\Delta_1\Pi_{xx} + \Delta_2\Pi_{yy} + \epsilon_1(t)\Pi_{xz} -    (\Gamma_{\phi 1} + \Gamma_{\phi 2})\Pi_{yx}\\
\dot{\Pi}_{xz}	& = & 2g\Pi_{0y} - \epsilon_2(t)\Pi_{xy} + \Delta_1\Pi_{yz} -  (\Gamma_{\phi 1}+\Gamma_{2})\Pi_{xz}\\
\dot{\Pi}_{zx}	& = & 2g\Pi_{y0} - \epsilon_1(t)\Pi_{yx} + \Delta_2\Pi_{zy} -  (\Gamma_{\phi 2}+\Gamma_{1})\Pi_{zx}\\
& & \\
\dot{\Pi}_{yy}	& = & -\Delta_1\Pi_{xy} - \Delta_2\Pi_{yx} + \epsilon_2(t)\Pi_{yz} + \epsilon_1(t)\Pi_{zy} -  (\Gamma_{\phi 1} + \Gamma_{\phi 2})\Pi_{yy}\\
& & \\
\dot{\Pi}_{yz}	& = & - \Delta_1\Pi_{xz} - \epsilon_2(t)\Pi_{yy} + \epsilon_1(t)\Pi_{zz} -  (\Gamma_{\phi 1}+\Gamma_{2})\Pi_{yz}\\
\dot{\Pi}_{zy}	& = & - \Delta_2\Pi_{zx} - \epsilon_1(t)\Pi_{yy} + \epsilon_2(t)\Pi_{zz} -  (\Gamma_{1}+\Gamma_{\phi 2})\Pi_{zy}\\
& & \\
\dot{\Pi}_{zz}	& = & -\epsilon_1(t)\Pi_{yz} -\epsilon_2(t)\Pi_{zy}  -  (\Gamma_{1} + \Gamma_{2})(\Pi_{zz}-Z_{T1}Z_{T2})
\end{array}
	\label{poxy}
\end{eqnarray}
Here we used the standard approximation for the dissipation operator $\hat{\Gamma}$ via the dephasing and relaxation rates to characterize the intrinsic noise in the system. Also, hereafter we assume for simplicity that relaxation rates are the same for both identical qubits, i.e.,  $\Gamma_{\phi 1} = \Gamma_{\phi 2}=\Gamma_{\phi}$ and $ \Gamma_{r1}=\Gamma_{r2}=\Gamma_{r}$, and that the temperature is low enough, resulting in the equilibrium values of the diagonal elements of the qubit density matrices being $ Z_{T2} = Z_{T1} = 1$.  All the simplifying assumptions (e.g.,
$\Delta_1=\Delta_2$, $\Gamma_{r1}=\Gamma_{r2}$, $\Gamma_{\phi 1}=\Gamma_{\phi 2}$ etc.)  do not 
qualitatively affect our results reported below. For instance, if $\Delta_1\ne \Delta_2$, the spectrum in Fig. 2 will have more peaks, corresponding
to larger numbers of levels due to the lifting of the artificial degeneracy.

In the limit of zero coupling $(g = 0)$, there exists a solution of Eqs.~(\ref{poxy}) with no entanglement between qubits. This solution can be written as a direct product of two single-qubit density matrices written through the corresponding Bloch vectors: $\hat{\rho}_j = \frac{1}{2} (1+X_j\hat{\tau}_x+Y_j\hat{\tau}_y+Z_j\hat{\tau}_z)$. The components of what can be called the Bloch tensor $\Pi_{ab}$ are then all zero except for $(\Pi_{ox},\Pi_{oy},\Pi_{oz})= (X_1,Y_1,Z_1)$ and $(\Pi_{xo},\Pi_{yo},\Pi_{zo})=(X_2,Y_2,Z_2)$. If the interaction is not zero, the entanglement between these qubits generates all the components of the Bloch tensor to be non-zero \cite{Zhang,i3} and such an entangled state persists  on the time scale
$1/\Gamma$ after  the interaction is later switched off [$g(t>t_0)=0$].   
  
This reflects the fact that, in the presence of interactions, the eigenstates of the system are entangled \cite{Zhang,i3}, and the noise terms in the eigenbasis will thus maintain the off-diagonal terms in the density matrix of the two-qubit system.

\subsection{Two flux qubits}

As a specific example of our approach, which can be experimentally implemented, we propose to measure two (almost) identical superconducting flux qubits consisting of a superconducting loop interrupted by four Josephson junctions and coupled via a coupler loop \cite{coupl} (See Figure 1). The state of each qubit is controlled by the applied magnetic  flux $\Phi_e^{(j)} = f_e^{(j)}\Phi_0$ through the loop, where $\Phi_0$ is the flux quantum. In the vicinity of $f_e^{(1)} = f_e^{(2)} = 1/2$, the ground state of the system is a symmetric superposition of the states $|L\rangle$ and $|R\rangle$, with a clock- and counterclockwise circulating superconducting current $I_p$, respectively. In the basis $\left\{ |L\rangle, |R\rangle\right\}$ the two-qubit system can be described by the
 Hamiltonian (\ref{eq-ham}) with $\epsilon_j = I_p\Phi_0\delta\! f_e^{(j)}$ with classical flux bias fluctuations $\delta\! f_e $ in the qubit loops around 1/2, while the tunneling amplitude $\Delta$ is determined by the fabrication of the loop and the junctions. Note that the components of the density matrix can be measured
{\em directly}, e.g., by monitoring the current fluctuations in the flux qubits: $I_{j}(t) = I_p X(t)$. The direct relation of this spectrum to the current/voltage noise spectrum in the resonant  tank $(LC)$ circuit coupled to the qubit was
used in Ref.~\onlinecite{i1}. 

\section{Simulation results}

Using the dimensionless time $\bar{t} = t \Delta$, we numerically solved the system (\ref{poxy}) by the Ito
method for two coupled qubits driven only by white classical noise, choosing parameters for damping
$\Gamma_{\phi}/\Delta=\Gamma_{r}/\Delta=0.1$ close to the ones experimentally found in flux qubits.
The spectra of $X_1=\Pi_{ox}$ and $Z_1=\Pi_{oz}$, for $g=0.5$, are shown in Fig.~2. Since this two-qubit system is only driven by noise, the spectrum of both $X_1$ and $Z_1$ is enhanced by increasing the noise, and peaks become more distinguished if noise is not too high. These spectra exhibit four maxima, whose positions nicely agree with the frequencies of the interlevel transitions (in   units of $\Delta$):
\begin{equation}
2\pi\nu_1=\omega_1=2g,\  \  2\pi\nu_{2,3}=\omega_{2,3}=\sqrt{1+g^2}\pm g,\  \ 2\pi\nu_4=\omega_4=2\sqrt{1+g^2}  
\end{equation}  
which have values $\nu_1\approx 0.16,\  \nu_2\approx 0.1, \nu_3\approx 0.26,$ and $\nu_4\approx 0.36$.
Two peaks out of these four frequencies are clearly seen on the $S_X$ spectra in Fig.~2, while the other two peaks are better seen on the $S_Z$ spectra. Either of these two spectra is sufficient to {\em measure both the coupling constant $g$ and  the tunneling splitting energy $\Delta$}, while the remaining spectrum can be used for control. Note here that, unlike the single-qubit case \cite{om}, there are peaks on both $S_X$ and $S_Z$ even for small values of the coupling strength $g$, which illustrates our earlier remark on the entangled nature of the eigenstates revealed by classical noise. 

To determine whether the coupling is ``ferro-" or ``antiferromagnetic", that is, the sign of the coupling constant $g$, we study the time correlations in the density matrix elements $ \Pi_{ox}(t)= X_1(t)$ and $ \Pi_{xo}(t)= X_2(t)$, for $g=\pm 0.7$ and $g=0$.  Numerically solving equations (\ref{poxy}) we obtained the time sequences $X_{j}(t_i)$   shown in Fig. 3, where $t_i$ is the discretized time of the simulation. Correlations and anticorrelations are clearly seen for ferromagnetic and antiferromagnetic coupled qubits, while almost no correlations are seen for the decoupled ones. 
 A qualitative physical picture of these correlations in the time domain is readily understood. For instance, for ferromagnetic coupling, the Bloch vectors of the two qubits tend to allign for weak enough noise. A stronger noise excites partially-coherent oscillations in the intervals between two sequential
noise spikes, but the qubit-qubit oscillations still tend to preserve the ferromagnetic ordering (which results in the correlations seen in Fig. 3) even for the dynamicaly evolving qubits. Similarly, the antiferromagnetic coupling tends to produce {\em anti}-correlations in the qubit dynamics, as seen in Fig.~3.

To quantitatively describe these correlations we plot the  sample Pearson correlation coefficient
\begin{equation}
r=\frac{n\sum_i\Pi_{ox}(t_i)\Pi_{xo}(t_i)-\sum_i\Pi_{ox}(t_i)
\sum_i\Pi_{xo}(t_i)}{ \sqrt{n\sum_i\Pi_{xo}^2(t_i)-\left( \sum_i\Pi_{xo}(t_i)\right)^2} \sqrt{n\sum_i\Pi_{ox}^2(t_i)-\left( \sum_i\Pi_{ox}(t_i)\right)^2} }
\end{equation}  
 as a function of the coupling constant $g$ (bottom panel of Fig.~3). Here   $n$ is the total number of simulation time steps.  
 The module of the correlation coefficient $r$ exhibits a maximum at $|g|\approx 0.7$. At larger $|g|$ the oscillations become weaker, since the
uncorrelated external noises in two qubits suppress each other via their coupling, so that the noise-induced oscillations weaken. The sign of $r$ coincides with the sign of the coupling $g$, which allows to easily distinguish between ferro- and antiferromagnetic couplings.

\section{Conclusions}

We have demonstrated that quantum correlations in a two-qubit system can be highlighted by the presence of classical noise. As an application of this effect, we suggest the use of noise spectroscopy. Namely, the measurement of the fluctuation spectra of the system, as a means to determine the relevant parameters of the multiqubit system. 

\section{acknowledgments}

We acknowledge partial support from the National Security
Agency, Laboratory of Physical Sciences, Army Research Office, National Science
Foundation (Grant No. 0726909), JSPS-RFBR (Grant No.
06-02-92114), MEXT Kakenhi on Quantum Cybernetics, FIRST (Finding Program for innovative R\&D on S\&T), and FRSF (Grant No. F28.21019), and  EPSRC
(No. EP/D072518/1). 

\begin{figure}[btp]
\begin{center}
\includegraphics[width=8.0cm]{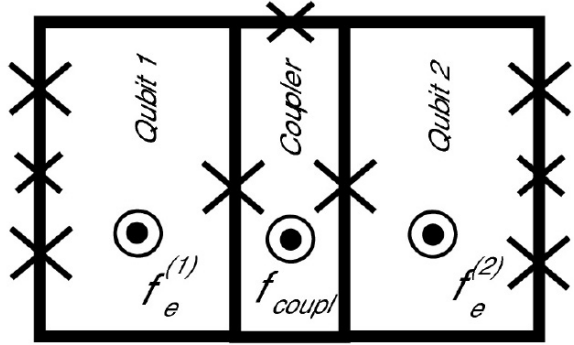}
\end{center}
\caption{Schematic diagram of two coupled flux qubits, each one with four Josephson junctions. 
These qubits can be coupled \cite{coupl} via the central coupler loop allowing to change the magnitude and sign of the coupling constant $g$.}
\end{figure}

\begin{figure}[btp]
\begin{center}
\includegraphics[width=10.0cm]{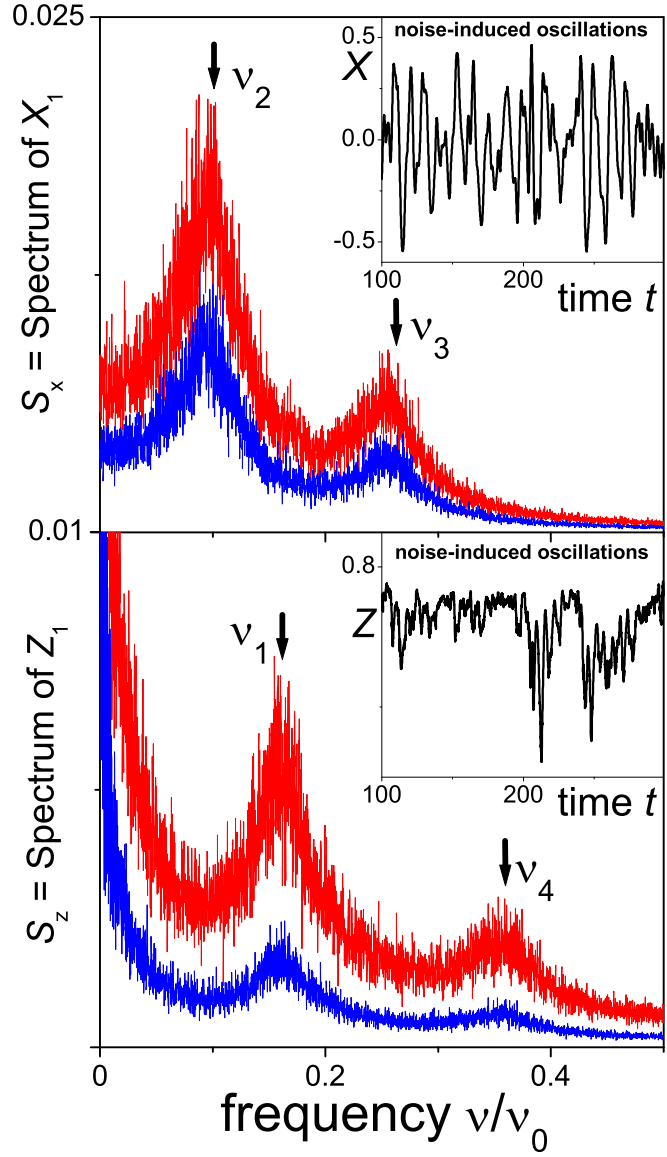}
\end{center}
\caption{(Color online.) Spectral density $S_X(\omega)$ (top panel) and $S_Z(\omega)$ for two values of the noise
($D/\Delta=0.04$ and $D/\Delta=0.013$) and normalized coupling $g/\Delta=0.5$. 
Four peaks, two per panel, can be easily distinguished, and these correspond to the four interlevel frequencies 
 $\nu_1, \nu_2, \nu_3,$ and $ \nu_4$. The insets show their corresponding time sequences $X(t)$ and $Z(t)$.}
\end{figure}

\begin{figure}[btp]
\begin{center}
\includegraphics*[width=10.0cm]{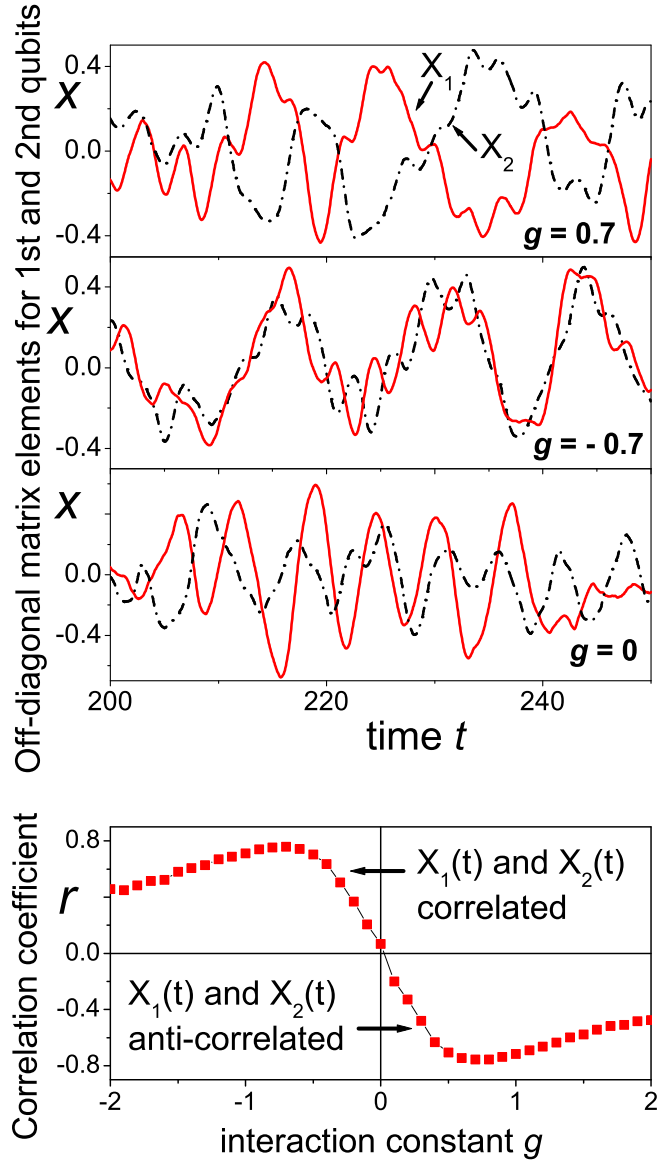}
\end{center}
\caption{ (Color online.) Time sequences for $X_1(t)=\Pi_{ox}(t)$ (continuous red curve) and $X_2(t)=\Pi_{xo}(t)$ (dot-dashed black curve) for values of the coupling constant $g=\pm 0.7$ (top two panels) and $0$ (third panel). The anticorrelations (top panel, $g>0$) and correlations (second panel, $g<0$) are clearly seen for nonzero coupling ($|g|=0.7$), while there are no correlations for $g=0$. The bottom panel shows the dependence of the correlation coefficient $r$ on the coupling constant $g$, when $D/\Delta=0.013$.}
\end{figure}


\begin{thebibliography}{99}
\bibitem{app}  The more standard form of the two qubit Hamiltonian is $H=-(1/2)\sum_{j=1,2} [\epsilon_j\sigma_z+\Delta_j\sigma_x^j]+g\sigma_x^1\sigma_x^2$. This Hamiltonian can be reduced to (\ref{eq-ham}) by the rotation
$\sigma_x\rightarrow\sigma_z$ and $\sigma_z\rightarrow \sigma_x$. In the paper we use Hamiltonian (\ref{eq-ham}) to simplify further 
notations.
\end{thebibliography}
\end{document}